\documentclass[11pt,3p,review,authoryear]{elsarticle}

\journal{International Journal of Forecasting}
\biboptions{longnamesfirst}

\usepackage{todonotes}
\usepackage{enumitem}
\usepackage{amsmath}
\usepackage{amsfonts}
\usepackage{bm}
\usepackage{url}
\usepackage{listings}

\urldef\github\url{https://github.com/M4Competition/M4-methods/tree/master/237%20-%20prologistica}

\begin{document}

\begin{frontmatter}

\title{Weighted Ensemble of Statistical Models}

\author[prologistica]{Maciej Pawlikowski\corref{cor}}
\ead{maciej.pawlikowski@prologistica.pl}
\author[prologistica]{Agata Chorowska}
\address[prologistica]{ProLogistica Soft\\ Ostrowskiego 9\\ 53-238 Wroclaw, Poland}
\cortext[cor]{Corresponding author}

\begin{abstract}
We present a detailed description of our submission for the M4 forecasting competition, in which it ranked
3\textsuperscript{rd}
overall.
Our solution utilizes several commonly used statistical models, which are weighted according to their performance on historical data.
We cluster series within each type of frequency with respect to the existence of trend and seasonality.
Every class of series is assigned a different set of models to combine.
Combination weights are chosen separately for each series.
We conduct experiments with a holdout set to manually pick pools of models that perform best for a given series type, as well as to choose the combination approaches.
\end{abstract}


\end{frontmatter}

\section{Introduction}\label{introduction}

Combining forecasts has been shown to greatly improve the forecast quality \citep{clemen}.
Averaging predictions produced by different models usually outperforms the individual methods.
Our submission for the M4 competition relies heavily on this technique.
We take several commonly used statistical models and weight their outputs according to their performance on a holdout set.
The main challenge was in choosing a pool of algorithms and a combination approach.

We categorize M4 data with regard to the frequency (monthly, weekly, etc.), as well as the existence of trend and seasonality.
For every category, we select a distinct pool of models.
A slightly different way of calculating weights is used for every frequency.

Section~\ref{description} is an overview of our combination methodology. Section~\ref{computationssummary} briefly summarizes the computations. Section~\ref{specialcases} describes additional heuristics that helped us further improve the accuracy.
Sections~\ref{discussion} and \ref{conclusion} provide the discussion and the conclusion.
In the Appendix we illustrate the complete forecast calculation process for an example time series.

\section{Method overview}\label{description}

\noindent Our forecasting method is comprised of five steps:
\begin{enumerate}
    \item Clustering time series.
    \item Choosing the model pool for each cluster.
    \item Measuring performance of the models with rolling origin evaluation.
    \item Determining weights for models in the pool.
    \item Calculating the final forecast.
\end{enumerate}

The method is parameterized by the model pools, the number of origins $N$ and the error averaging formula $f$ used in the rolling origin evaluation, as well as the model weighting formula $g$.
The choice of each parameter is performed separately.
For a given cluster, the model pool is chosen first, then the rolling origin evaluation parameters, and finally, the weighting formula.
Each series is split into the training part and the holdout part, where the length of the holdout part is equal to the respective forecast horizon.
We monitor the Overall Weighted Average (OWA) error \citep{m4guide} on the holdout part to choose the best parameter values.
The OWA was the main accuracy measure in the M4 competition.

\subsection{Clustering time series}
We group all series into classes according to the frequency and the existence of seasonality and trend.

To detect seasonality we use a 90\% autocorrelation test similar to the one used in M4 benchmarks, while
to detect trend we employ a Mann-Kendall test \citep{gilbert} with 95\% confidence.

Trend and seasonality detection are omitted for daily series during clustering because we did not find it beneficial. Splitting daily series into multiple classes did not help us lower the holdout set error.
We also do not use it for the frequencies with low representation in the dataset (hourly and weekly) in order to avoid possible overfitting of model pools when the classes get too small.

Since the yearly series are not seasonal, this means we had 13 classes in total: 2 for yearly series, 4 for both monthly and quarterly, and a single class for each of weekly, daily, and hourly data.

\subsection{Choosing the model pool}
Each class of series is assigned a distinct model pool. The list of individual models we choose from, alongside respective R packages and functions, is given below. To calculate forecasts we used R 3.5.0 with packages "forecast" (version 8.2) \citep{forecastr} and "forecTheta" (version 2.2) \citep{forectheta}.

\begin{itemize}
    \item \textit{Na\"ive models} - Na\"ive 1, Na\"ive 2 \citep{m4guide}

    \item \textit{Exponential Smoothing (ETS) models} - simple ETS [{forecast::ses}], ETS with automatic model choice [{forecast::ets}], with or without damped trend \citep{HKSG02}

    \item \textit{Theta models} - Theta method \citep{theta} [{forecast::thetaf}], Optimized Theta method \citep{otm} [{forecTheta::otm}]

    \item \textit{ARIMA models} - ARIMA with automatic parameter choice [{forecast::auto.arima}]

    \item \textit{Linear regression models} - linear regression of a series on various types of a trend (constant, linear, logarithmic) [{stats::lm}], optionally applied to deseasonalized data
\end{itemize}

Seasonality in automatic ETS and ARIMA methods is handled by their implementations in the "forecast" package.
In other cases we use classical multiplicative decomposition.

These models form a default model pool for all clusters.
The final pool for each cluster is a subset of this default pool.
There is a slight exception to this rule for hourly, daily and weekly data, where multiple variants of each model may be present in the default pool.
For hourly data, we consider two seasonality periods (24 and 168) by including two variants of each model.
Each variant assumes a different seasonality period.
For weekly and daily data, we add variants of models that operate on trimmed series, which allows us to focus on the latest observations.
In the case of weekly data, we trim the series to the last few years.
For daily series, we leave the last several weeks.
The final pool may contain multiple variants of the same model and all of them may receive non-zero weights.
Additionally, for hourly, daily, and weekly data we consider only trimmed variants of the ARIMA and ETS models to reduce the computational complexity.
More details on this step can be found in the source code of our submission\footnote{\github} and in the Appendix.

\subsubsection*{Model pool selection procedure}
\noindent
To determine the model pool for a class of time series, we follow the procedure below:
\begin{enumerate}
    \item
    Start with a default pool, set the rolling origin evaluation parameters and the weighting formula to default values (described in later sections).
    \item For each series in the class:
        \begin{enumerate}
            \item Split the series into a training part and a holdout part.
            \item For each model, forecast the series with that model. Calculate the holdout set error.
            \item Determine the combination weight for each model (sections \ref{rollingorigin}, \ref{choosingweights}).
        \end{enumerate}
    \item Sort the models by their mean holdout set error across all series in the class.
    \item Combine all models. Compute the mean holdout set error of the combination.
    \item Try removing models from the pool starting with the one with the highest holdout set error:
        \begin{enumerate}
            \item Remove the model. Compute the mean holdout set error of the combination.
            \item If it increases, put the model back into the pool. Then, move to the next model.
            \item Stop when all models have been considered.
        \end{enumerate}
\end{enumerate}

\subsection{Rolling origin evaluation}\label{rollingorigin}

For each model in the pool and each series in the class, perform a rolling origin evaluation \citep{tashman} with a constant window size 1.
The number of origins $N$ depends only on the series' frequency, thus, for example, all monthly series share the same value.
For each series, we average the symmetric Mean Absolute Percentage Error (sMAPE) across origins using a weighting function~$f$. This produces a vector of performance scores of models that is later converted into the combination weights (described in section \ref{choosingweights}). We choose~$f$ from among a simple arithmetic mean and a weighted mean with exponential weights, in which the later origins count more. The default value of~$f$ is a simple mean. Similar to $N$, $f$ is chosen per frequency.

We find the best $N$ and $f$ after the model pool choice. For $N$, we check all values up to the respective forecast horizon (the default value of $N$) and monitor the holdout set errors of the combination. Then we check all possible values of $f$, holding $N$ fixed. It is worth noting that this means the rolling origin evaluation is performed twice for a given series: once on the training part, during parameter tuning, and once on the whole series, to calculate weights for the final prediction.

Table~\ref{tabN} shows the chosen values of $N$. In most cases, we were able to make it significantly lower than forecast horizon without observing any meaningful difference in error magnitude on the holdout set. This is desirable because the computational cost of our method scales linearly with $N$.

\begin{table}[ht]
\centering
\begin{tabular}{|l|c|c|c|c|c|c|}
\hline
\textbf{Frequency} & yearly & quarterly & monthly & weekly & daily & hourly \\ \hline
\textbf{$\bm{N}$ / $\bm{h}$}      & 3 / 6    & 8 / 8        & 10 / 18     & 13 / 13    & 8 / 14    & 24 / 48     \\ \hline
\end{tabular}
\caption{Chosen values of $N$ alongside respective forecast horizons $h$ for all frequencies.}
\label{tabN}
\end{table}

For the $f$ function, the default mean aggregation worked well for most frequencies.
The two exceptions were daily and yearly series, where we observed a slight improvement using exponential weights.

We experimented with three error measures in the rolling origin evaluation: the sMAPE, the Mean Squared Error (MSE) and the OWA.
We observed no noticeable difference in OWA error on the holdout set between these metrics.
Since the choice of a metric for this step didn't seem to affect the accuracy of the combination, we decided to use the sMAPE as it was the most convenient.

\subsection{Determining combination weights}\label{choosingweights}
The sMAPE errors are converted into weights using a formula $g$ taking one of the following forms. A small epsilon is added to the denominator to avoid division by zero.
\[
    \begin{split}
    g_{inv} (S) &= 1\ /\ (S + \epsilon) \\
    g_{sqr} (S) &= {g_{inv}}^2 (S) \\
    g_{exp} (S) &=  \exp(g_{inv}(S))
    \end{split}
\]
$S$ is a vector of performance scores for a given series calculated in the rolling origin evaluation step.
All operations above are element-wise (i.e. applied separately to each element of $S$).
The formula $g$ is chosen per frequency, using exhaustive search while monitoring the mean holdout set error of the combination.
The default value of $g$ is $g_{sqr}$.
$g_{inv}$ has been chosen for hourly series and $g_{exp}$ for weekly series. In other cases the default $g_{sqr}$ performed the best, possibly because $g$ was fixed to a default value during the model pool fitting phase.

The graphs displaying the mean model weights for each frequency can be found in the Appendix.

\subsection{Calculating final forecast}
Calculate forecasts of the series for models in the pool.
The final prediction is defined as a weighted mean of those forecasts, using weights described in section~\ref{choosingweights}.
Moreover, since the M4 dataset contains no negative values, any negative forecasts are replaced with zeros.

\section{Computations summary}\label{computationssummary}

Here we provide a brief summary of the computations described in the previous section. Let $X$ be a yearly time series of length 30 with horizon 6.
For the sake of simplicity, let's assume we chose $N$ = 3, $f = $ arithmetic mean, $g = g_{sqr}$, and a set of models $m_1, \dots, m_5$.

\begin{enumerate}
    \item
    Perform the rolling origin evaluation:
    \begin{enumerate}
        \item
        Calculate $e_{i,j}$, where $i \in \{1,2,3,4,5\}$, $j \in \{1,2,3\}$, and $e_{i,j}$ is the sMAPE error of the one-step-ahead forecast produced by $m_i$ fitted to the first $30-j$ terms of $X$
        \item
        Average the evaluation results for each model:
        $s_i = (e_{i,1} + e_{i,2} + e_{i,3})\ /\ 3$
    \end{enumerate}

    \item
    Compute the model weights: $w_i = s_i^{-2}$

    \item
    Calculate $F_1, \dots, F_5$ -- 6-steps-ahead forecasts created with $m_1, \dots, m_5$ fitted to $X$

    \item
    The final forecast is a weighted average of $F_1, \dots, F_5$ using weights $w_1, \dots, w_5$
\end{enumerate}

\section{Special cases}\label{specialcases}

In addition to the above algorithm, we employed two heuristic procedures that improved the accuracy on the holdout set. These two additional steps occur after computing combination forecasts described in Sections~\ref{description} and \ref{computationssummary}.

\subsection{Daily series}

In the case of daily series, we were unable to find a model pool that would significantly outperform a single Na\"ive 1 model.
Thus, we use forecast combinations for only a part of daily data.
Since the Na\"ive 1 predictor is best suited for time series originating in a random walk, we have heuristically identified such cases.
For each daily series, we compare the sMAPE error on the holdout set of the Na\"ive 1 forecast to a threshold $t_{rnd}$.
If the error is smaller, we label the series as \textit{random} and forecast it exclusively with Na\"ive 1.

We chose the value of $t_{rnd}$ after choosing the model pool, $N$, $f$, and $g$.
We fixed those parameters and tested 10 evenly spaced $t_{rnd}$ values from the interval $\left[0.01, 0.1\right]$ (the threshold had to be small).
For each value, we determined which series are considered \textit{random}, replaced the combination forecast for these series with the Na\"ive 1 prediction, and computed the mean holdout set error for all daily series.
During experiments we settled on a threshold of 0.05, which resulted in roughly 90\% of daily series being considered \textit{random}.

\subsection{Forecast by analogy}

We were able to use forecasting by analogy to significantly boost the holdout set accuracy for daily and hourly data.
We use the correlation coefficient to determine which series should be predicted this way.
The correlation is computed between the last values of a given series and every window of every series in the dataset.
Properly scaled and shifted continuation of the most correlated window replaces the combination forecast for the series, provided the correlation is strong enough.
The detailed process for a series $X = (x_1, \dots, x_n)$ is described below.
The procedure is parameterized by the window length $M$ and the threshold $t_{cor}$.
The forecast horizon for $X$ is denoted by $h$. $\overline{X}$ denotes the mean of the series and $\sigma(X)$ is the standard deviation.

\begin{enumerate}
    \item Let $X' = (x_{n-M+1}, \dots, x_{n})$ be the last $M$ observations of $X$.

    \item For every series $Y = (y_1, \dots, y_m)$ with the same frequency as $X$:
    \begin{enumerate}
        \item For $i$ from 1 to $m-h-M+1$, compute the correlation coefficient between the window $(y_i, \dots, y_{i+M-1})$ and $X'$.
    \end{enumerate}

    \item Let $Z = (z_1, \dots, z_k)$ denote the series containing the window most correlated with $X'$. Let $Z' = (z_j, \dots, z_{j+M-1})$ be that window. If the correlation exceeds a threshold $t_{cor}$:
    \begin{enumerate}
        \item Let $Z'' = (z_{j+M}, \dots, z_{j+M+h-1})$ be the continuation of $Z'$ with length $h$.
        \item Replace the forecast for $X$ with $(Z'' - \overline{Z'}) \dfrac{\sigma(X')}{\sigma(Z')} + \overline{X'}$.
    \end{enumerate}
\end{enumerate}

We arbitrarily picked $M = 2h$.
The value of $t_{cor}$ is chosen as the last parameter of our method. For each frequency, we manually tested several values from $\left[0.95, 0.999 \right]$.
For each value, we computed the holdout set error averaged across all series with the given frequency.
We achieved the largest accuracy boost for 0.99 on daily data and 0.995 on hourly data. These values resulted in 33\% of daily series and 40\% of hourly series being predicted by analogy.
For yearly and weekly series the results did not improve when using this method.
We did not have enough time to try forecasting by analogy on monthly and quarterly data before the competition ended, but we did so after the test data has been released.
For quarterly series, we did not observe any improvement.
However, for monthly series this method decreased the mean sMAPE error on the test set from 12.747 to 12.624.
With $t_{cor}$ set to 0.995, 10\% of monthly series were predicted by analogy.

\section{Discussion}\label{discussion}

Combining statistical models was a very popular approach in the M4 competition \citep{m4results}.
We believe that the key to success of our method was the careful choice of model pools and weights.

The choice of model pool was crucial in our experiments.
Averaging all models never turned out optimal.
We also observed that simply combining several top performing models did not result in the best choice either.
For example, including the Na\"ive 1 model in the pool in many cases improved the accuracy, even though as a single model it often performed the worst.

After the test data has been released, we investigated the impact of the averaging function $f$ and the weighting formula $g$ on the three largest datasets: yearly, quarterly, and monthly.
Different choices of $f$ changed the test set error in a meaningful way only for monthly series.
On the other hand, altering the formula $g$ had a major impact on the accuracy in all cases.

In the case of series predicted by analogy, we did not use model combinations.
It is possible that combining forecasting by analogy with other models would yield a better result.

We decided to use the window size 1 during the rolling origin evaluation step.
This choice has been made due to its simplicity.
We also tested the evaluation using one window with size $N$ instead of $N$ windows with size 1, but this method was significantly less accurate.
There may still, however, exist a more optimal way to obtain performance scores.

The tuning of per-cluster parameters has been performed manually.
We fit each of the parameters independently, in order to make it feasible given the time constraints.
While manual inspection can provide intuitions about the impact of particular variables on the final performance, a proper grid search should ultimately result in a more optimal set of values.
It would also enable lower-level choices, like using different values of $N$ for different classes within the same frequency, which becomes physically impossible without automation.

Finally, the difference in the holdout set accuracy between the full list of models and the chosen pool was the largest for frequencies with a small amount of series (hourly and weekly).
This might mean that a more fine-grained clustering of data could help improve the forecast quality.
Considering more time series characteristics would result in smaller and more homogeneous clusters, which should make it easier to fit specialized model pools.

\section{Conclusion}\label{conclusion}
In this paper, we described the forecasting methodology used in the M4 competition.
The core of our approach is the combination of statistical models, with rolling origin evaluation determining the weights.
The most important and the hardest task was choosing which models to combine.
We found the choice of the model pool to have a major impact on the accuracy, sometimes in an unexpected way.
Including even weak models in the combination might improve the overall performance.
At the same time, we found the fitting of the pool to be necessary, as the default setting was never optimal.
We believe this should be the main takeaway from this research.

\section*{Acknowledgements}


The authors would like to thank the developers of the "forecast" package for R, which has been used extensively in this research.

Funding:
This work was supported by the European Regional Development Fund [grant number RPDS.01.02.01-02].

\bibliography{refs}

\end{document}






\section*{Appendix}

\subsection*{\textbf{Numerical forecast calculation example}}

\begin{figure}[!htb]
    \centering
    \includegraphics[width=\textwidth]{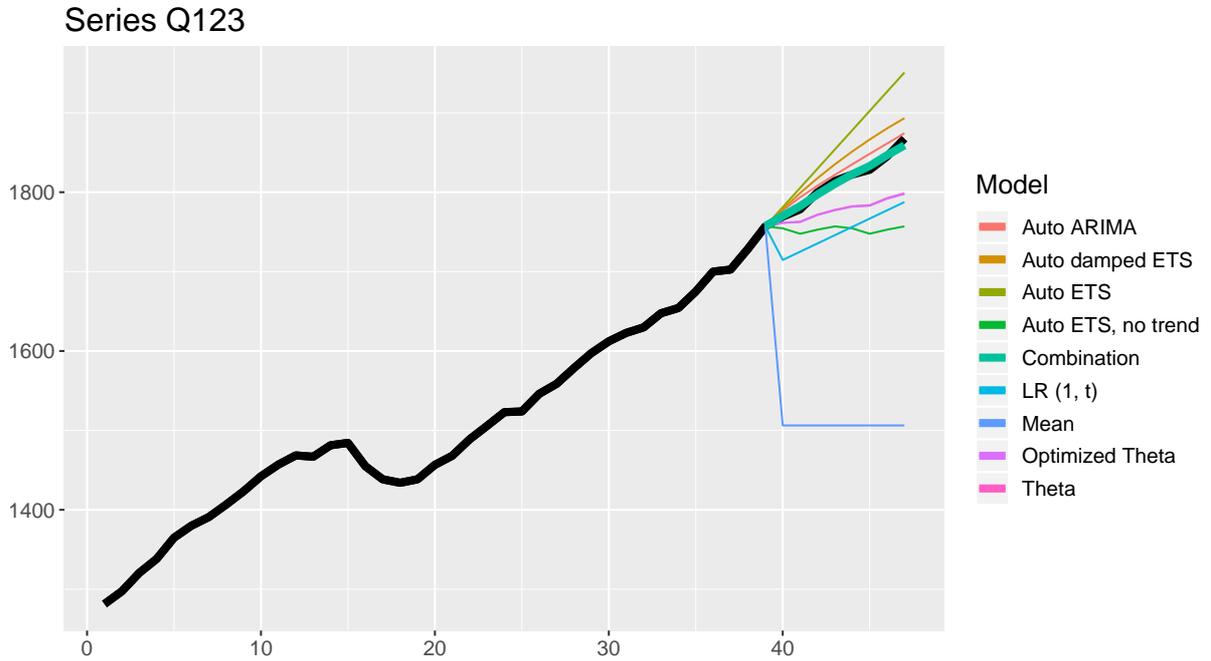}
    \caption{Forecasts of models in the pool and their combination for quarterly series Q123. Historical and future values of a series are drawn with a black line. Since the Theta and Optimized Theta forecasts were almost identical in this case, the difference between them is not visible in the graph. Naming conventions: Auto - automatic model choice, LR (1, t) - linear regression with two trend regressors (constant and linear), Mean - mean value of the series, damped - allow trend dampening, no trend - disable trend component.}
    \label{q123plot}
\end{figure}

In this section we present the forecast calculation process for a single quarterly series Q123 (Table~\ref{q123numbers}).
Series Q123 has been classified as series with trend but with no seasonality. 
The model pool for that class has been comprised of 8 models, as shown in Figure~\ref{q123plot}. 
For quarterly series, the chosen number of origins for the rolling origin evaluation step was 8. 
The errors were averaged across origins using an arithmetic mean. 
The weighting formula was an inverse square.

Table~\ref{q123rolling} contains the rolling origin evaluation results for the classic Theta method. 
The number of origins is 8, so the model is fitted 8 times. 
The series value at origin $k$ is the $k^{\mathrm{th}}$ from last value of the series.
The forecast at origin $k$ is created with a model fitted to the series without the last $k$ values. 
The weight for the Theta method is calculated as an inverse square of the mean sMAPE error listed in Table~\ref{q123rolling}: $0.007345856^{-2} = 18531.7$. 

We compute the rolling origin evaluation values and the combination weights in a similar way for the rest of the models in the pool. 
Table~\ref{q123weights} lists the weights for all models. 
The predictions of the individual models are then combined using a weighted average to create a final forecast. 
This is displayed in Table~\ref{q123forecast} and Figure~\ref{q123plot}.

\begin{table}[!htb]
\centering
\small
\begin{tabular}{lrrrr}
              & \textbf{Quarter 1} & \textbf{Quarter 2} & \textbf{Quarter 3} & \textbf{Quarter 4} \\
\textbf{2005} & 1281.37            & 1297.41            & 1320.54            & 1338.16            \\
\textbf{2006} & 1364.89            & 1379.98            & 1390.85            & 1406.64            \\
\textbf{2007} & 1423.32            & 1442.23            & 1456.97            & 1468.53            \\
\textbf{2008} & 1466.84            & 1481.30            & 1484.30            & 1454.99            \\
\textbf{2009} & 1438.39            & 1434.04            & 1438.41            & 1456.65            \\
\textbf{2010} & 1468.11            & 1488.86            & 1505.77            & 1523.02            \\
\textbf{2011} & 1523.84            & 1546.09            & 1558.71            & 1578.53            \\
\textbf{2012} & 1597.39            & 1612.19            & 1622.79            & 1629.73            \\
\textbf{2013} & 1647.54            & 1654.14            & 1674.93            & 1699.99            \\
\textbf{2014} & 1702.52            & 1728.56            & 1756.94            &                   
\end{tabular}
\caption{Quarterly series Q123 from the M4 dataset.}
\label{q123numbers}
\end{table}

\begin{table}[!htb]
\centering
\small
\begin{tabular}{lrrrrrrrr}
\textbf{Origin}       & 8       & 7       & 6       & 5       & 4       & 3       & 2       & 1       \\
\textbf{Series value} & 1629.73 & 1647.54 & 1654.14 & 1674.93 & 1699.99 & 1702.52 & 1728.56 & 1756.94 \\
\textbf{Forecast}     & 1627.34 & 1634.35 & 1652.24 & 1658.90 & 1679.77 & 1704.93 & 1707.53 & 1734.47 \\
\textbf{sMAPE}        & 0.0129  & 0.0122  & 0.0014  & 0.0120  & 0.0096  & 0.0011  & 0.0080  & 0.0015 \\[5pt]
\textbf{Mean sMAPE}        & \multicolumn{2}{r}{0.007345856}
\end{tabular}
\caption{Rolling origin evaluation results for the Theta method on the series Q123. For each origin, the model is fitted to the training data, and a one-step-ahead forecast is created. The table lists series values at each origin alongside the forecast values, the sMAPE error between the two, and the sMAPE error averaged across origins.}
\label{q123rolling}
\end{table}

\begin{table}[!htb]
\centering
\small
\begin{tabular}{lrr}
\textbf{}                   & \textbf{Weights} & \textbf{Normalized weights} \\
\textbf{Auto ARIMA}         & 24117.0          & 0.2022                      \\
\textbf{Auto damped ETS}    & 22189.2          & 0.1860                      \\
\textbf{Auto ETS}           & 24503.9          & 0.2054                      \\
\textbf{Auto ETS, no trend} & 9607.0           & 0.0805                      \\
\textbf{LR (1, t)}          & 2138.4           & 0.0179                      \\
\textbf{Mean}               & 58.4             & 0.0005                      \\
\textbf{Optimized Theta}    & 18133.6          & 0.1520                      \\
\textbf{Theta}              & 18531.7          & 0.1554                     
\end{tabular}
\caption{The combination weights for the series Q123. The normalized weights sum up to 1.}
\label{q123weights}
\end{table}

\begin{table}[!htb]
\centering
\scriptsize
\begin{tabular}{lrrrrrrrrr}
                            & 1       & 2       & 3       & 4       & 5       & 6       & 7       & 8       & \textbf{sMAPE}  \\
\textbf{Auto ARIMA}         & 1777.17 & 1793.55 & 1808.12 & 1821.84 & 1835.15 & 1848.27 & 1861.30 & 1874.29 & 0.0065          \\
\textbf{Auto damped ETS}    & 1778.71 & 1798.92 & 1817.70 & 1835.13 & 1851.32 & 1866.35 & 1880.30 & 1893.26 & 0.0134          \\
\textbf{Auto ETS}           & 1781.17 & 1805.40 & 1829.63 & 1853.87 & 1878.10 & 1902.33 & 1926.57 & 1950.80 & 0.0271          \\
\textbf{Auto ETS, no trend} & 1754.48 & 1747.64 & 1752.83 & 1756.93 & 1754.48 & 1747.64 & 1752.83 & 1756.93 & 0.0349          \\
\textbf{LR (1, t)}          & 1714.64 & 1725.06 & 1735.48 & 1745.89 & 1756.31 & 1766.73 & 1777.15 & 1787.57 & 0.0361          \\
\textbf{Mean}               & 1506.29 & 1506.29 & 1506.29 & 1506.29 & 1506.29 & 1506.29 & 1506.29 & 1506.29 & 0.1861          \\
\textbf{Optimized Theta}    & 1761.21 & 1762.44 & 1771.28 & 1777.25 & 1781.74 & 1782.92 & 1791.81 & 1797.78 & 0.0206          \\
\textbf{Theta}              & 1761.51 & 1762.81 & 1771.73 & 1777.77 & 1782.34 & 1783.59 & 1792.56 & 1798.61 & 0.0203          \\[5pt]
\textbf{Combination}        & 1770.34 & 1782.41 & 1797.16 & 1810.52 & 1822.59 & 1833.02 & 1846.58 & 1858.96 & \textbf{0.0018} \\[5pt]
\textbf{Future values}      & 1769.22 & 1778.36 & 1799.83 & 1814.19 & 1822.28 & 1828.16 & 1845.01 & 1867.53 &                
\end{tabular}
\caption{Forecasts of the hidden future values created with individual models and their combination. Future values are shown in the bottom row. The rightmost column lists the sMAPE errors of the forecasts.}
\label{q123forecast}
\end{table}

\clearpage
\subsection*{\textbf{Mean model weights per frequency}}

\begin{figure}[!htb]
    \centering
    \includegraphics[width=\textwidth]{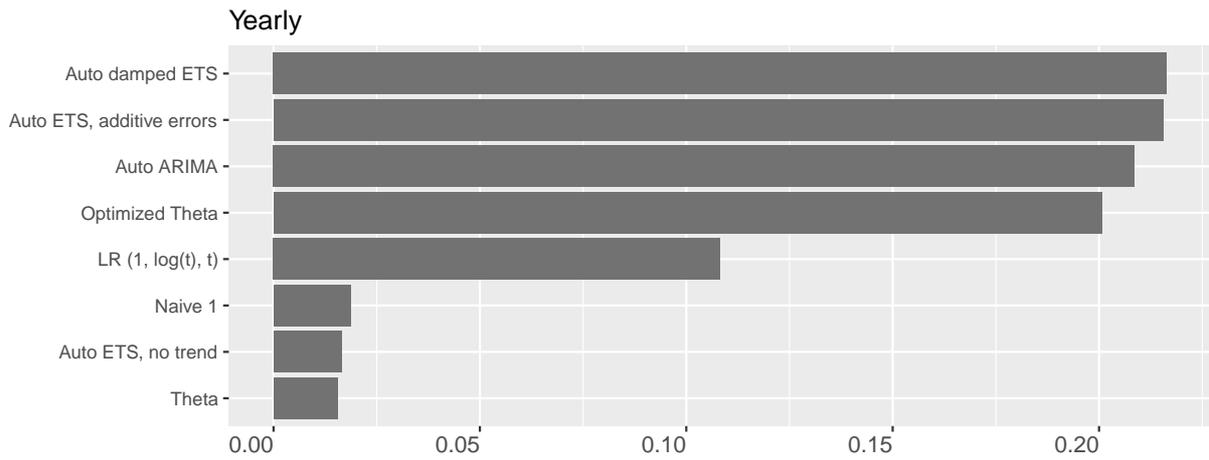}
    \caption{Mean normalized model weights for yearly series.}
    \label{weightsy}
\end{figure}

\begin{figure}[!htb]
    \centering
    \includegraphics[width=\textwidth]{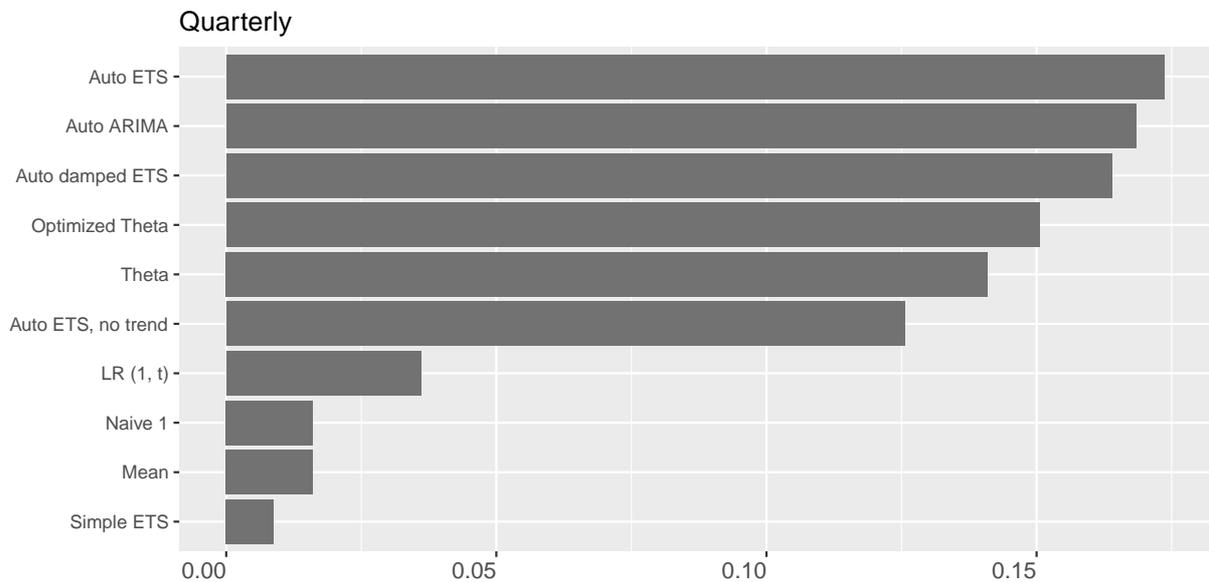}
    \caption{Mean normalized model weights for quarterly series.}
    \label{weightsq}
\end{figure}

\begin{figure}[!htb]
    \centering
    \includegraphics[width=\textwidth]{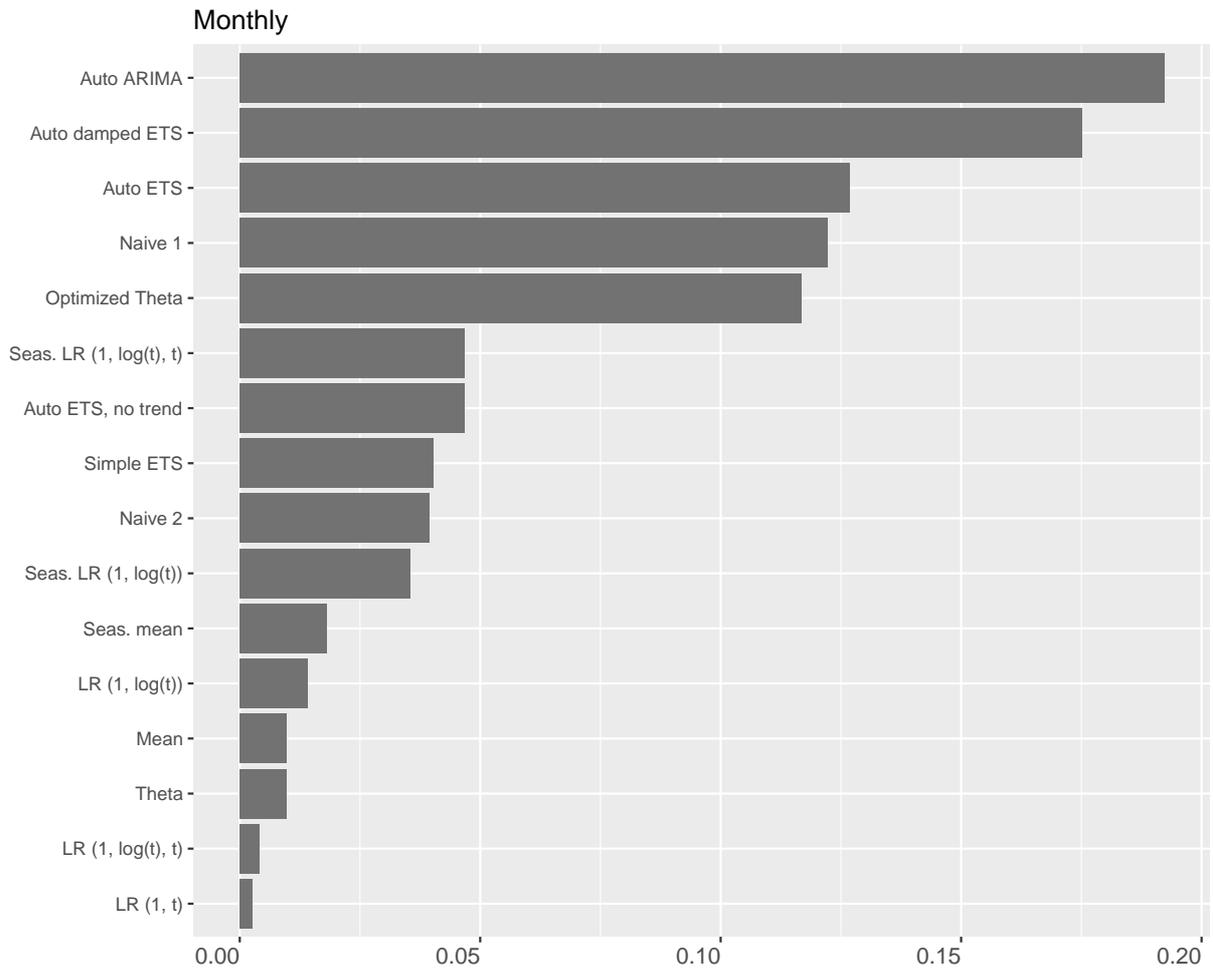}
    \caption{Mean normalized model weights for monthly series. Prefix "Seas." indicates the model is applied to deseasonalized data.}
    \label{weightsm}
\end{figure}

\begin{figure}[!htb]
    \centering
    \includegraphics[width=\textwidth]{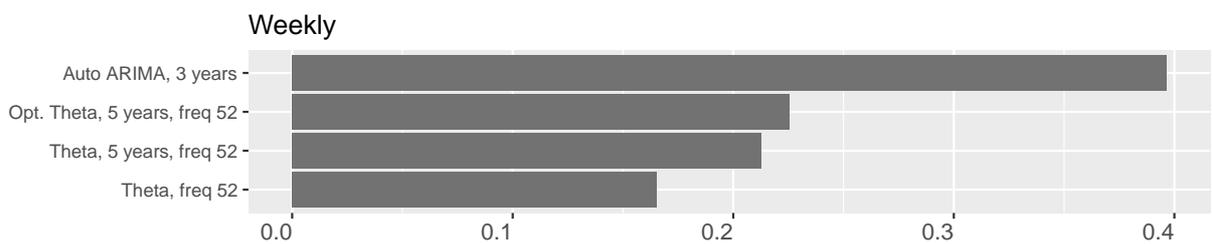}
    \caption{Mean normalized model weights for weekly series. If the time period is specified, the model is applied to a trimmed series. If the frequency is specified, the seasonality detection assumes this frequency. For example "Opt. Theta, 5 years, freq 52" is the Optimized Theta method fitted to the last 5 years of the series, assuming the seasonality period length 52.}
    \label{weightsw}
\end{figure}

\begin{figure}[!htb]
    \centering
    \includegraphics[width=\textwidth]{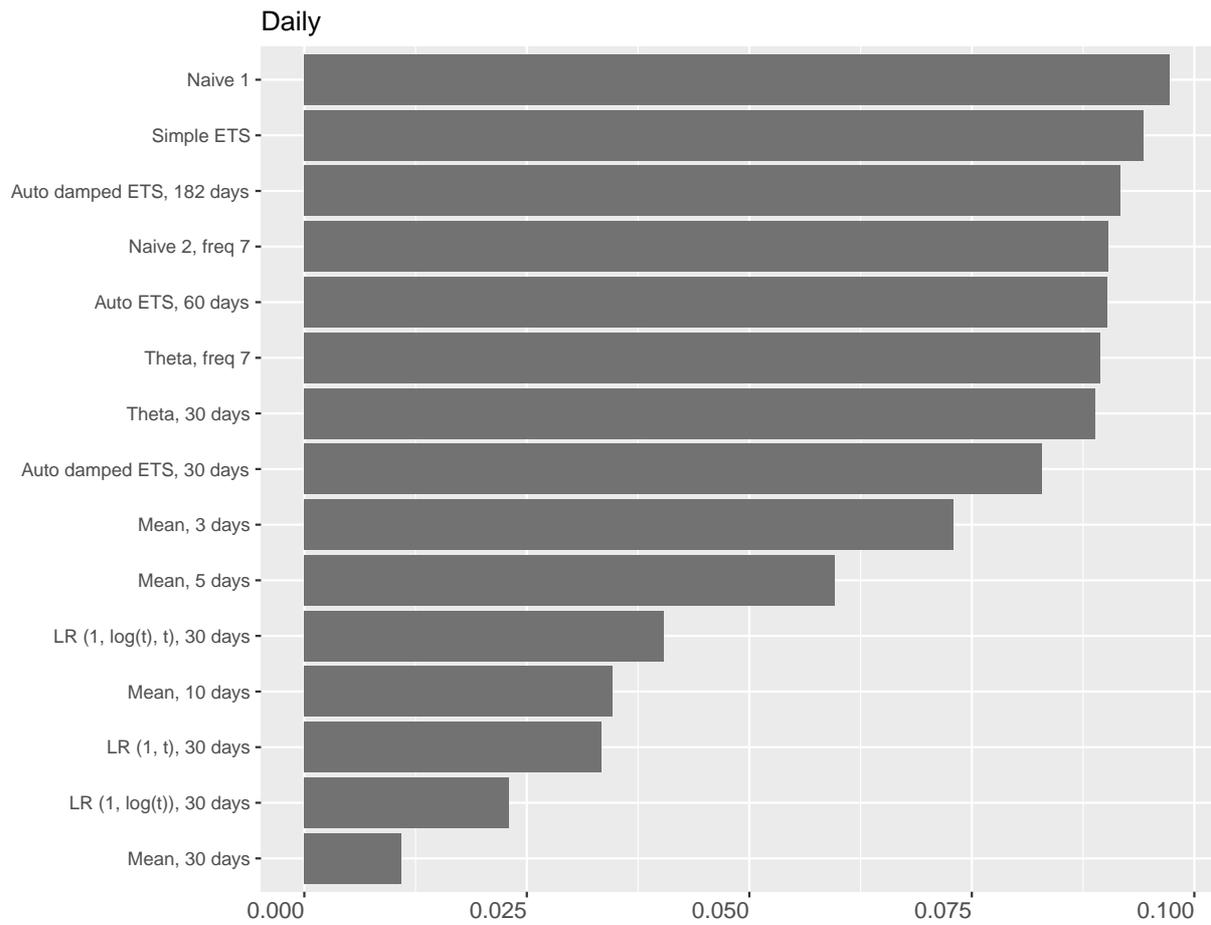}
    \caption{Mean normalized model weights for daily series. The weights here pertain only to 6\% of all daily series (the ones predicted with combination). Other daily series were predicted with Na\"ive 1 (61\%) or by analogy (33\%).}
    \label{weightsd}
\end{figure}

\begin{figure}[!htb]
    \centering
    \includegraphics[width=\textwidth]{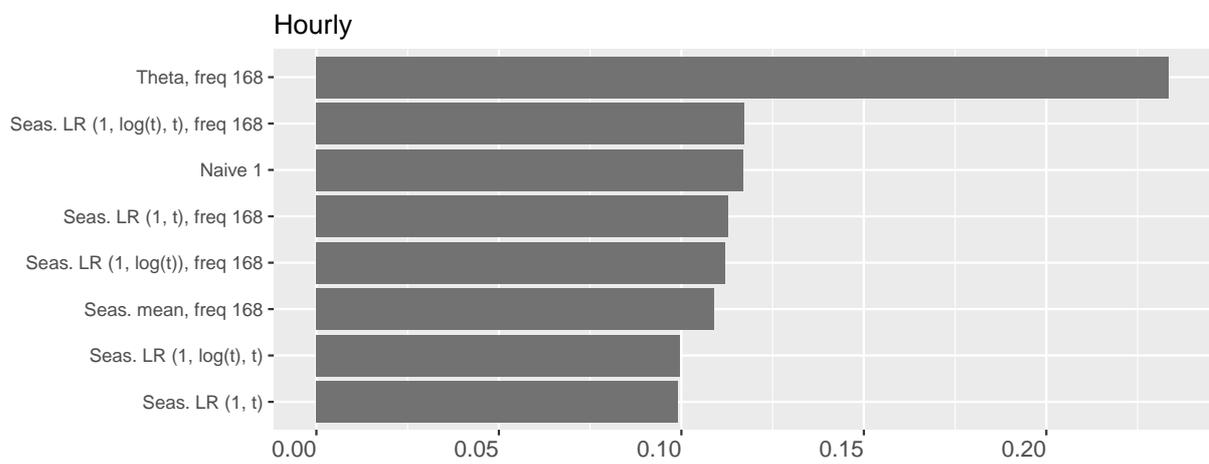}
    \caption{Mean normalized model weights for hourly series. 60\% of hourly series were predicted with combination, the rest by analogy.}
    \label{weightsh}
\end{figure}

In this part, we show the mean normalized model weights for each frequency. Note that there were usually multiple clusters within each frequency. 
Thus, the information presented in Figures \ref{weightsy} to \ref{weightsh} is a summary of all model pools for a given frequency. 
Model naming convention is the same as in Figure~\ref{q123plot}.

